\documentclass[runningheads,letterpaper]{eptcs}
\usepackage{amssymb}
\usepackage{amsmath}
\usepackage{verbatim}
\usepackage{graphicx}
\usepackage{xspace}
\usepackage{multicol}
\usepackage{url}

\begin{document}

\title{How Computers Work:\\Computational Thinking for Everyone}

\author{
Rex Page\thanks{This material is based upon work supported by the National Science Foundation under Grant No. 1016532.}  \\
University of Oklahoma \\
Norman, OK, USA \\
\and
Ruben Gamboa \\
University of Wyoming \\
Laramie, WY, USA}

\def\authorrunning{Rex Page and Ruben Gamboa}
\def\titlerunning{How Computers Work: Computational Thinking for Everyone}

\maketitle

%\markboth{Rex Page and Ruben Gamboa}
%         {How Computers Work: Computational Thinking for Everyone}

\begin{abstract}
What would you teach if you had only one course to
help students grasp the essence of computation
and perhaps inspire a few of them to make computing a subject of further study?
Assume they have the standard college prep background.
This would include basic algebra, but not necessarily more advanced mathematics.
They would have written a few term papers, but would not have written computer programs.
They could surf and twitter, but could not exclusive-or and nand.
What about computers would interest them or help them place their experience in context?
This paper provides one possible answer to this question by discussing
a course that has completed its second iteration.
Grounded in classical logic, elucidated in digital circuits and computer software,
it expands into areas such as CPU components and massive databases.
The course has succeeded in garnering the enthusiastic attention of
students with a broad range of interests,
exercising their problem solving skills, and
introducing them to computational thinking.
\end{abstract}

\section{One and Done}
\label{sec:one-and-done}

What would you teach if you had only one\footnote{Exceptional athletes
in American colleges sometimes enroll with the intention of
turning professional after one year.
Coaching vernacular for these athletes is ``one and done.''
The title of this section applies the term to students
who may take only one course in computation.}
course to
help students grasp the essence of computation and perhaps
inspire a few of them to make computing a subject of further study?
Assume they have the standard college prep background.
This would include basic algebra, but not necessarily more advanced mathematics.
They would have written a few term papers, but would not have written computer programs.
They could surf and twitter, but could not exclusive-or and nand.
What about computers would interest them or help them place their experience in context?

This paper discusses one of the many possible answers to this question.
It describes experiences in teaching an honors course for students from a
variety of disciplines at the University of Oklahoma.
The students have varied interests and come from all college levels,
first year to fourth year.
They can choose from many courses to satisfy their honors requirements,
from Beatles History to Moby Dick to What is Science?
This course, called ``How Computers Work: Logic in Action,''
has succeeded in getting the enthusiastic attention of some of these students and
in exercising their problem solving skills.

The course includes some computer programming, but does not dwell on it.
Students get enough experience to know what software is, but not
enough to take on serious software development projects.
The material helps students understand what makes automated computation
possible by expanding on computational principles and overarching insights
rather than the details needed in the practice of engineering.
Most of the students will not continue with additional study in computer science.
They will not become practicing engineers in hardware or software development.

Computers are demystified.
Students grasp the fundamentals that make automated computation possible.
We have some objective support for these claims from student assessments of the course
and from scores on examinations. Exam scores averaged over 90\% in both offerings
of the course, and we think the exams call for comparable problem solving skills
and insight about concepts than other exams
we have given in computer science courses over the years.

Student assessments (which are discussed in more detail in Section \ref{sec:assessment})
are mostly, but not uniformly, positive about the understanding
of the workings of computers that they have acquired through their studies in the course.
One student was disappointed that the course failed to discuss
``the actual physical mechanisms of computers and how the different parts
(e.g. RAM, motherboard, etc.) work together.''
Others were similarly surprised by the content of the course,
but they were happy rather than disappointed with what they learned.
Based on the overall tenor of student assessments,
we think students did acquire a perspective
on computers that make them less mysterious.
Our conclusion is guided more by our past experience in teaching
computer science than in objective evidence.

The honors course is an outgrowth of a course in applied logic
required in the computer science program at the University of Oklahoma.
Other papers have discussed the earlier course and supporting tools
\cite{PageCompLogicUG,EastlundDracula,PageFPTPforUG,VaillancourtACL2DrS,PageSWisDM,PageFPWhere,PageESC}.

The honors course includes more material on the big picture of computational thinking
and more writing projects (as opposed to problem solving projects)
than the original logic course,
but there is enough overlap to make it an adequate substitute, at least for good students.
Topics from the required course that are omitted in the honors version to make room
for more coverage of the big picture include Karnaugh maps, quicksort, and a few other algorithms.
Coverage of deductive reasoning is reduced in the honors course, but
equation-based reasoning is covered at about the
same level as in the required course.
These changes make it possible
to include material on large applications such as Google's use of MapReduce \cite{mapreduce}
for rapid searches
and Facebook's Cassandra approach to the massive database problem \cite{cassandra}.

The material gives students who are not majoring in computer science
a leg up, but does not overlap directly with mainstream material
in most computer science programs.
Such programs, at least in the United States, usually discuss logic in
a required course in discrete mathematics, but the material on logic
rarely takes up more than 20\% of the discrete mathematics course
and usually does not discuss directly in any detail connections between
logic and the design of computer hardware and software.

The big ideas in the honors course are
\begin{enumerate}
\item{the correspondence between digital circuits and formulas in logic,}
\item{how abstractions facilitate combining solutions to small problems to form solutions to big ones,}
\item{how algebraic formulas can specify computations,}
\item{how models expressed in software capture the behavior of processes and devices,}
\item{how important, complex algorithms derive from simple, definitional properties,}
\item{how different definitional properties can produce the same results at vastly different computational expense,}
\item{how computational expense makes some useful devices feasible and renders others infeasible,}
\item{and
how all of these ideas bear on the ability of computers to deal with information
on the massive scale needed to provide services like search engines,
internet storefronts, and social networks.}
\end{enumerate}

The idea of deriving new properties by rigorous reasoning from definitional ones
is suffused throughout the course.
Algebraic equations specify software and digital circuits as formulas in logic, and
logic is employed to derive new equations. In this sense, logic forms the basis
of both the objects under study and analysis of those objects.
These ideas have been around for a long time.
John McCarthy wrote about them extensively over a half-century ago \cite{McCarthy}.\footnote{In a sense
the course originates from in-person interactions with McCarthy.
He visited the University of Oklahoma for two days
in 1996 at the invitation of the School of Geology and Geophysics to speak about
the sustainability of human progress.
The School was attracted to McCarthy by some material on his website
that was favorable to their interest in expanding exploration for oil.
The geophysicists did not feel that they could productively entertain McCarthy
for two days, so they asked the director of the School of Computer Science
(which happened to be one of the authors)
to escort McCarthy for meals and to arrange
for him to deliver a separate seminar for the School of Computer Science.
It was a thrilling opportunity.

At breakfast one morning,
the subject of introducing an undergraduate course
on reasoning about software arose.
McCarthy remarked that logic would be the essence of the course.
This remark provided the inspiration for organizing the course with logic as the main
theme and with reasoning about software providing examples of logic in action.
It was truly an ``aha'' moment leading, albeit by a long and circuitous route,
to the course that this paper describes.}

Based on our observations of the understanding of the big ideas by most of
the computer science students we encounter, we think that the honors students
that decide to go further have a better initial understanding of the big ideas
than many of the students who decide early on to focus their studies in computing.
One student commented that
``As someone who was already acquainted with a traditional programming language,
I thoroughly enjoyed learning the methods and ideas presented in this class.''
Another said the course ``helped me understand how computers
communicate through their languages and axioms.''
We do not think many incoming computer science students would mention
axioms as an element of their understanding of computers.
Several made comments indicating that the course unlocked new parts of their brains,
again an unusual take on a computing course compared to that of most computer science students.

%%\cite{ref-label}

\section{Demographics}
\label{sec:demographics}

``How Computers Work: Logic in Action''
has been offered twice, so far, as part of a collection of
``perspectives courses'' in the Honors College at the University of Oklahoma.
As one of many requirements for earning a degree with honors,
students must complete two perspectives courses.
The first offering of this perspectives course was in spring, 2011.
By popular demand, a repeat offering took place in spring, 2012,
and  the course is scheduled again for 2013.

Perspectives courses in the Honors College are limited to nineteen students.\footnote{The
reason for the limit of nineteen, rather than twenty or some other number,
has been lost. It was decided over fifteen years ago, and no member of the
current Honors College faculty or staff has been with the College that long.}
Nineteen students enrolled in the 2011 offering, but two dropped
the course after a few weeks.
The 2012 offering was oversubscribed at twenty students.
One dropped, leaving a full contingent of 19 to complete the course.

Eight of the total of 36 students in the two offerings of the course
had major fields of study outside
science and engineering:
history, letters, philosophy, linguistics, economics, drama, psychology, and business.
Sixteen were majoring in science, eleven were engineering students,
and one, a first-year student, had not yet decided on a major.
One of the engineering students was in computer science,
and three were in computer engineering.
Engineering students scored, on the average,
three percentage points higher on examinations than students in science
and five points higher than students outside science and engineering.

Almost 80\% of the students (28 of 36) were in their first two years of college.
About 60\% of the students had some prior experience in programming.
In most cases this was a course in high-school or college, but
five students had been programming for more than two years.
None had any prior experience in functional programming.
Students with prior programming experience scored,
on the average, four percentage points higher on examinations
than students without programming experience.

All students in the course are honors students, which means they held
a grade-point average of at least 3.4 (out of a possible 4.0)
at the time of joining the honors program and would need to
maintain at least that average to have an honors designation
(``cum laude'') on their diploma at graduation.
In addition, enrolling in honors courses suggests a
high level of self motivation.
Honors students tend to be well-engaged in their studies,
and they participate energetically in class.
They ask interesting questions,
and vague or sloppy answers seldom go unremarked.

To summarize, about 20\% of students enrolled in the course were
majoring in humanities, social sciences, or business.
About 30\% were engineering majors
(mechanical, electrical, chemical, computer engineering, computer science),
and about 45\% were science majors
(physics, chemistry, biochemistry, microbiology, meteorology).\footnote{The
percentages do not add up to 100 because one student had not yet chosen a major.}
Two of the science students were simultaneously working on a degree in mathematics.
Almost 80\% of the students were in their first two years of college.
Engineering students scored higher, on the average, on examinations
than students in arts and sciences, but not by much.
Among arts and science students, science students did marginally better,
but the margin was only two percentage points.

\section{Assessment}
\label{sec:assessment}

Near the end of the course, the university asks students to complete a 20-question
assessment of their experience in each of their courses.
The full record of student responses to these questions
for both offerings of the course
is accessible online \cite{studevals}.
We provide a summary in this section that we think gives an accurate picture of the
range of student opinion.
About 70\% of the students completed the questionnaire in both offerings of the course.
The questionnaire has fifteen questions requiring a ranking on a five-point scale
(5 for ``far above average'' down to 1 for ``far below average'')
and five questions calling for free-form comments.

Assessments of both offerings of the course were mostly positive.
The ratings in 2012 were a bit higher than in 2011.
For example, the median responses to the questions
``How intellectually stimulating was this course?'' and
``How much did this course help you develop your critical thinking skills?''
were 4 in 2011, but 5 in 2012.
In both years the median rating was 5 for
``Overall this course was $\langle r \rangle$,''
but the mean went from 4.75 in 2011 to 4.9 in 2012.

Responses to the question ``What were the specific strong points of the course?''
included observations about
interesting, novel concepts and readings,
the quality of posted lecture notes,
and improvements
in critical thinking coming from
challenging concepts and homework.
One student responding to a question about the weak points of the course
said that the rigor of the course was tedious at times.
Another said that the overall goal of the course was unclear.
Other comments pointed out that the concepts were difficult for people not familiar with computers,
that there was too much jargon,
and that explanations in the textbook  were sometimes incomplete or unsatisfactory.
Some students complained that the course did not cover as many
big concepts and ideas or have as many in-class discussions as other perspectives courses
in the Honors College.

Another question asked for an overall opinion of the course.
Most responses to the question were positive:
``This course was beyond excellent.''
``It was a lot of work, and very difficult, but I enjoyed it and am glad I took it.''
The course ``encourages difficult, critical thinking.''
There were three negative responses to the question.
One student asked for more writing assignments
to make the course more like other perspectives courses,
and two students wanted more material on the ``big picture.''

The Honors College administration was pleased enough
with student reactions in both
years to ask for another offering in 2013.

\section{Course Content}
\label{sec:content}

\subsection{Equations}
\label{sec:equations}

Much of the material in the course centers around equations.
For example, software is introduced as operations that transform
operands to results. This is the standard viewpoint of functional programming.
Of course any software (or hardware), transforms input signals
to output signals, which makes it possible to take the functional viewpoint
regardless of paradigm.
However, this is only a high-level picture, and the details
will differ substantially in different paradigms.

In this course, all
software is described in the form of equations.
That is, all programs in the course are functional programs.
This is partly to keep the amount of material within reasonable bounds
for a one-semester course, but mostly to simplify reasoning about
software and digital circuits and to make it possible to use a mechanized
logic to formalize some of the proofs.

Operations are described informally, through templates that relate
inputs to results.
For example we define the list constructor, {\tt cons},
with the following informal equation.
\begin{samepage}
\label{cons-def}
\begin{center}
\begin{tabular}{lll}
({\tt cons} $x$ $[x_1 ~ x_2 ~ ... ~ x_n]$) = $[x ~ x_1 ~ x_2 ~ ... ~ x_n]$ & ~~~~~~~~~~ & \{\emph{cons}\}
\end{tabular}
\end{center}
\end{samepage}

The traditional definition of {\tt cons} would avoid informality
by introduction the constructor and its destructors together
and avoid the need for names of elements of the list:
\begin{samepage}
\label{cons-def}
\begin{center}
\begin{tabular}{lll}
({\tt first} ({\tt cons} $x$ $xs$)) = $x$  & ~~~~~~~~~~ & \{\emph{fst-id formal}\}\\
({\tt rest~} ({\tt cons} $x$ $xs$)) = $xs$ & ~~~~~~~~~~ & \{\emph{rst-id formal}\}
\end{tabular}
\end{center}
\end{samepage}

We think it works better to start with a less formal approach
and work towards greater formality.
That is why our initial presentation of {\tt cons} has a
list template with a name for each list element.

Informally, we use square brackets to delimit templates for lists
and $n$ to denote an arbitrary natural number.
In this case (and most cases) subscripts
distinguish the names of individual elements in the list.
In the above example $[x_1 ~ x_2 ~ ... ~ x_n]$ stands for a list with $n$ elements,
and $[x ~ x_1 ~ x_2 ~ ... ~ x_n]$ is a list with $n+1$ elements whose first element is $x$
and whose subsequent elements are $x_1$, $x_2$, and so on up to $x_n$.

We put labels for equations on the right in curly braces.
Later, we refer to an equation by its label.
We would refer to the foregoing equations as
\{\emph{cons}\}, \{{\emph{fst-id formal}\}, and \{{\emph{rst-id formal}\}.

To be consistent, we use informal list templates
in our initial definitions of {\tt first} and {\tt rest},
as with did with {\tt cons}, instead of the more formal presentation in
\{{\emph{fst-id formal}\} and \{{\emph{rst-id formal}\}.
\begin{samepage}
\label{first-rest-defs}
\begin{center}
\begin{tabular}{lll}
({\tt first} $[x_1 ~ x_2 ~ ... ~ x_{n+1}]$) = $x_1$                  & ~~~~~~~~~~ & \{\emph{fst}\}\\
({\tt rest~} $[x_1 ~ x_2 ~ ... ~ x_{n+1}]$) = $[x_2 ~ ... ~ x_{n+1}]$ & ~~~~~~~~~~ & \{\emph{rst}\}
\end{tabular}
\end{center}
\end{samepage}

We ask students to specify tests that the operations
would have to satisfy if they were functioning correctly.
The equations \{\emph{fst-id formal}\} and \{\emph{rst-id formal}\}
provide an example of tests of this kind.
Following our usual habits, we present the tests informally:
\begin{samepage}
\label{first-rest-defs}
\begin{center}
\begin{tabular}{lll}
({\tt first} ({\tt cons} $x ~ [x_1 ~ x_2 ~ ... ~ x_n])) = x$                      & ~~~~~~~~~~ & \{\emph{fst-id}\}\\
({\tt rest~} ({\tt cons} $x ~ [x_1 ~ x_2 ~ ... ~ x_n])) = [x_1 ~ x_2 ~ ... ~ x_n]$ & ~~~~~~~~~~ & \{\emph{rst-id}\}\\
\end{tabular}
\end{center}
\end{samepage}

Students have no difficulty convincing themselves that
if {\tt cons}, {\tt first}, and {\tt rest} fail to satisfy
either of the equations, \{\emph{fst-id}\} and \{\emph{rst-id}\},
there must be something wrong with at least one of the operators.

\subsection{Tests}
\label{sec:tests}

In tandem with informal templates, we introduce
a formal notation, DoubleCheck \cite{EastlundDracula},
to specify tests expressing our expectations about
the values that formulas stand for.\footnote{DoubleCheck
is similar to QuickCheck \cite{classenQuickCheck}, except that DoubleCheck
is embedded in the Dracula environment for ACL2 and QuickCheck is embedded
in Haskell. DoubleCheck does not have the sophisticated facility for narrowing
counterexamples that QuickCheck provides, which means the DoubleCheck would be
much less effective for industrial-strength usage, but we have found it to be effective
for the purposes of our course.}

This makes it possible for students to practice expressing their expectations in a form
that allows the computer system to perform tests automatically using random data.
\begin{verbatim}
  (defproperty fst-id
    (x  :value (random-integer)
     xs :value (random-list-of (random-integer)))
    (equal (first (cons x xs))
           x))
  (defproperty rst-id
    (x  :value (random-integer)
     xs :value (random-list-of (random-integer)))
    (equal (rest (cons x xs))
           xs))
\end{verbatim}

Some students complain about oddities in the notation:
``{\tt equal}'' for ``='',
prefix notation instead of infix,
fully parenthesized formulas instead of relying on rules of precedence, etc.
But, they accept these things as necessary for
communicating with the computing system.
We think one reason for uncomplaining acceptance
is that students are struggling with much more difficult
concepts, and details of notation provide a welcome retreat into the trivial. \footnote{Sometimes
students do not accept arbitrary concepts so easily.
For example, some students object strenuously to the truth table for the implication operator.
The table is derived from the basic equations, so their objection in this matter is more
philosophical than technical.
They think the truth table simply fails to capture the usual meaning of implication.
Our defense avoids philosophy altogether.
We point out that we are not using logic for discussing issues from everyday life.
We are using it to explain the inner workings of digital circuits and software.
The implication operation is useful in this enterprise, whether or not its
properties conform to normal expectations. Of course, we know that it does conform
to standard uses in natural language, but getting into a debate about that point
does not do much to advance the educational goals of the course.}

Complaints fade quickly.
There are plenty of more challenging things for the students to think about.
Besides, having multiple notations for the same mathematical object is
a theme that comes up again when students see the correspondence between
formulas in Boolean algebra and diagrams of digital circuits,
with decimal numerals and binary numerals as representations of
natural numbers, and in other examples.

DoubleCheck properties employ mutable state and higher order operations to generate
random values from type specifications.
This is impossible in ACL2
because ACL2 does not support those language characteristics,
but Dracula \cite{VaillancourtACL2DrS} runs property-based tests
in the Racket \cite{racket} environment
(the successor to DrScheme \cite{DrScheme}),
which does support them.
We do not explain the higher-order nature of DoubleCheck properties in
the course.
Based on student reactions, it seems unlikely
that adding such a discussion would bring more clarity to the material.\footnote{More recently
we have been using a programming environment
called {\tt Proof} {\tt Pad} developed by Caleb Eggensperger \cite{proofpad}.
It lacks some of the facilities of Dracula,
especially a modules system for packaging and controlling the visibility of identifiers,
but is easier to install and runs faster.
{\tt Proof} {\tt Pad} does provide adequate facilities, including DoubleCheck,
for a course with logic as its central theme.}

\subsection{Inductive Definitions}
\label{sec:inductive-defs}

Inductive definitions appear painlessly in the context of the testing of expectations.
An early example is an operator {\tt append} for concatenating lists.
\begin{quote}
({\tt append}~$[x_1 ~ x_2 ~ ... ~ x_m] ~ [y_1 ~ y_2 ~ ... ~ y_n]$) = $[x_1 ~ x_2 ~ ... ~ x_m ~ y_1 ~ y_2 ~ ... ~ y_n]$
\end{quote}

Students recognize that if the operator {\tt append}
failed to pass either of the following tests,
\{\emph{app1}\} and \{\emph{app0}\},
it could not be functioning properly.
\begin{samepage}
\label{append-def}
\begin{center}
\begin{tabular}{lll}
({\tt append}~({\tt cons} $x ~ xs$)~$ys$) = ({\tt cons} $x$~({\tt append}~$xs ~ ys$)) & ~~~~~~ & \{\emph{app1}\}\\
({\tt append}~$[~] ~ ys$) = $ys$                                                      & ~~~~~~ & \{\emph{app0}\}
\end{tabular}
\end{center}
\end{samepage}

Initially, we avoid viewing these equations as an inductive definition.
They are billed as simple tests that a correctly functioning operator would pass.
Later, we assert that any such collection of equations actually
defines an operator, provided the equations have the following characteristics:
\begin{enumerate}
\item{\emph{Consistent}: no two equations specify different results for the same input;}
\item{\emph{Comprehensive}: all forms of input must match the operands
      on the left side of at least one equation;}
\item{\emph{Constructive}: any inductive reference to an operator must comprise a reduced computation.}
\end{enumerate}

Students learn that under these conditions,
all properties of the operator derive from the definitional equations.
In other words, the equations define the operator.
We refer to these characteristics as ``the three C's.''
They are recurring theme from early on and throughout the course.

We describe an inductive reference to an operator, {\tt op},
as an invocation of {\tt op} on the right-hand side of an equation, \{\emph{eqn}\},
that also refers to {\tt op} on the left-hand side.
The operands in the inductive reference
must match the operands on the left-hand side of a non-inductive
equation more closely than they match the operands
on the left-hand side of \{\emph{eqn}\}.
That is, the reference on the right-hand side of the equation
represents a reduced computation compared to the formula on the left-hand side.

We do not attempt a rigorous definition of the term ``reduced computation.''
We do not go into detail about the degree of matching between operands, either.
The reduced computation issue, especially, can be subtle, but is not subtle in any of the
inductive references used as examples in the course.
In many cases the operands in inductive references are
lists that are shorter than the corresponding operands
on the left-hand side of the equation,
and the corresponding operands in non-inductive equations
are of a fixed, shorter length, usually zero, sometimes one or two.
In other cases, an argument that is expected to be a natural number
is smaller in the inductive reference than on the left-hand side,
and non-inductive equations specify the corresponding operand
as a fixed, smaller value, usually zero.

Since students do not encounter in the course subtle reductions in the
size of the computation expressed by inductive references,
we think that including a definition of computation size or
degree of matching in operands would be more obfuscatory than helpful.
Similarly, we believe that a discussion of fixpoints would take
us far afield of what we are tying to communicate in the course.
Accordingly, we have not tried to include these ideas in the discussion.

However, offering students the opportunity to explore
these ideas in one or more of the writing projects they are required to
complete as part of their course work could bring substantial
rewards to capable students. Probably, one-on-one discussions
of the ideas during office hours would be a necessary form of guidance,
and that could be rewarding to both student and instructor.
We are intrigued by the notion
of going in this direction in future offerings of the course.

Another subtle issue has to do with matching in data structures
and substitution of new, equivalent formulas into parts of existing formulas.
We point out to students that this is a non-trivial activity and
that early explorations in formal logic got the definition of
matching and substitution wrong more than once before arriving at
a correct definition. In this case, too, we rely on students to
figure it out from practice rather from a formal definition,
which we believe
students would find more confusing than helpful.

We think there is some evidence for our position in this matter
in our discussion of the correspondence between digital circuit diagrams
and Boolean formulas. We do present a formal definition of the isomorphism,
but students routinely ignore it. They manage to work out
correspondences between formulas and circuit diagrams in practice without
referring to the formal definition. When they try to apply the formal definition,
they usually get lost in details and technicalities.

\subsection{Inductive Proofs}
\label{sec:inductive-proofs}

The {\tt append} operator is associative.
\begin{center}
({\tt append}~$xs$~({\tt append}~$ys ~ zs$)) = ({\tt append}~({\tt append}~$xs ~ ys$)~ $zs$)
\end{center}

Students express this property formally in the notation of the DoubleCheck facility.
They run the test and find that it succeeds.
\begin{verbatim}
  (defproperty app-assoc
    (xs :value (random-list-of (random-integer))
     ys :value (random-list-of (random-integer))
     zs :value (random-list-of (random-integer)))
    (equal (append xs (append ys zs))
           (append (append xs ys) zs)))
\end{verbatim}

Of course the associativity property, like all properties of the {\tt append} operator,
can be derived from its definitional properties, \{\emph{app1}\} and \{\emph{app0}\}.
We derive properties of operations mostly by substituting new, equivalent,
formulas at strategic points to form new equations, a method entirely
familiar from high-school algebra, except that the operators involved,
instead of being addition, multiplication and the like,
often deal with non-numeric data, such as lists.
Since the definitional equations are usually inductive,
most of our derivations cite an induction hypothesis
at some point to justify moving from one formula to another, equivalent one.

A pencil-and-paper proof of the associativity property
from the informal equations could be carried out
as an induction on the length of $xs$.
The base case, when $xs$ is the empty list,
cites the \{\emph{app0}\} equation twice.

\begin{samepage}
\label{app-assoc-base}
\begin{center}
\begin{tabular}{llll}
  & ({\tt append}~$[~]$~({\tt append}~$ys ~ zs$))   & ~~~~~~ &                \\
= & ({\tt append}~$ys ~ zs$))                       & ~~~~~~ & \{\emph{app0}\}\\
= & ({\tt append}~({\tt append}~$[~] ~ ys$) ~ $zs$) & ~~~~~~ & \{\emph{app0}\}
\end{tabular}
\end{center}
\end{samepage}

In the inductive case, the length of $xs$ is non-zero.
That is, $xs$ = $[x_1 ~ x_2 ~ ... ~ x_{n+1}]$ for some natural number $n$.
So, the inductive case can be argued as follows.

\begin{samepage}
\label{app-assoc-ind}
\begin{center}
\begin{tabular}{llll}
  & ({\tt append}~$[x_1 ~ x_2 ~ ... ~ x_{n+1}]$~({\tt append}~$ys ~ zs$))              & ~~~~~~ &                    \\
= & ({\tt append}~({\tt cons}~$x_1$~$[x_2 ~ ... ~ x_{n+1}]$)~({\tt append}~$ys ~zs$))  & ~~~~~~ & \{\emph{cons}\}    \\
= & ({\tt cons}~$x_1$~({\tt append}~$[x_2 ~ ... ~ x_{n+1}]$~({\tt append}~$ys ~zs$)))  & ~~~~~~ & \{\emph{app1}\}    \\
= & ({\tt cons}~$x_1$~({\tt append}~({\tt append}~$[x_2 ~ ... ~ x_{n+1}] ~ ys$)~$zs$)) & ~~~~~~ & \{\emph{ind hyp}\} \\
= & ({\tt append}~({\tt cons}~$x_1$~({\tt append}~$[x_2 ~ ... ~ x_{n+1}] ~ ys$)~$zs$)) & ~~~~~~ & \{\emph{app1}\}    \\
= & ({\tt append}~({\tt append}~({\tt cons}~$x_1 ~ [x_2 ~ ... ~ x_{n+1}]$)~$ys$)~$zs$) & ~~~~~~ & \{\emph{app1}\}    \\
= & ({\tt append}~({\tt append}~$[x_1 ~ x_2 ~ ... ~ x_{n+1}] ~ ys$)~$zs$)              & ~~~~~~ & \{\emph{cons}\}
\end{tabular}
\end{center}
\end{samepage}

The DoubleCheck property {\tt app-assoc} is a formal statement of associativity,
and students can direct Dracula to submit the property to
the ACL2 theorem prover \cite{KaufmannCAR} for formal, mechanized verification.
In this way, students gain experience with inductive definitions
and with expressing expectations both formally and informally.
The also learn to do informal, paper and pencil proofs, and
they see how a mechanized logic can be used to produce formal proofs.
This is an important point, philosophically, because students observe
that it is easy to get paper-and-pencil proofs wrong.
Formalization is necessary for assurance that claims
about properties of software or hardware are correct.

\subsection{Programming}
\label{sec:programming}

Suppose a student wants to define an operator
that extracts the first $n$ elements from a list $xs$.
Using the principle of the three C's described in Section \ref{sec:inductive-defs},
the student looks for equations that express properties
of the operator that are consistent, comprehensive, and constructive.
Students who manage to conjure up the following equations
have succeeded in writing a program for the {\tt prefix} operator.
\begin{samepage}
\label{cons-def}
\begin{center}
\begin{tabular}{lll}
({\tt prefix} $0 ~ xs$) = $[~]$                                                         & ~~~~~~~~~~ & \{\emph{pfx0}\}   \\
({\tt prefix} $n ~ [~]$) = $[~]$                                                        & ~~~~~~~~~~ & \{\emph{pfx}$-$\} \\
({\tt prefix} $(n+1)$ ({\tt cons} $x ~ xs)$) = ({\tt cons}~$x$ ({\tt prefix} $n ~ xs$)) & ~~~~~~~~~~ & \{\emph{pfx1}\}
\end{tabular}
\end{center}
\end{samepage}

We do not pretend that inventing these equations is easy.
It calls for creativity and insight, and those things come from practice.
Students need to do lots of exercises to learn the material.
Gradually, they build up their expertise,
and along with it comes the ability to derive
new properties of operators from definitional ones.
The entire mechanism is built on ordinary, algebraic equations
and classical logic.\footnote{One could
provide guidelines beyond the three C's, such as mimicking in function
definitions the patterns of induction found in data structure definitions,
following, for example, the pedagogy of the textbook \emph{How to Design Programs}
\cite{HtDP}.
However, we do not expect students to become accomplished programmers.
We want them to experience a few creative insights and
to understand how computers can interpret algebraic equations as programs.
We hope that a few of them may be inspired
to study software development more seriously.
At that point they can acquire a facility with design patterns that will
enable them to be good programmers.}

To run their program for the {\tt prefix} operator, they must formalize it
in ACL2 \cite{KaufmannCAR}.
Since there are three equations,
the definition will need to say which formula for ({\tt prefix} $n ~ xs$)
applies in what circumstances. It can use the ``{\tt if}'' operator
to make the appropriate selection.
In this example, even though there are three equations,
there are only two distinct formulas for the value of ({\tt prefix} $n ~ xs$)
because the empty list is the result that {\tt prefix} delivers in two of the equations.
So, the ``{\tt if}'' operator only needs to choose between two
formulas, and the definition could be written as follows.\footnote{ACL2
is based on Common Lisp. To run ACL2 programs,
the Dracula programming environment converts them to Scheme
and employs Racket \cite{racket} to interpret them.
However, Dracula uses the ACL2 mechanized logic to prove theorems.}

\begin{verbatim}
  (defun prefix (n xs)
    (if (and (consp xs) (not (zp n)))
        (cons (first xs) (prefix (- n 1) (rest xs)))  ; {pfx1}
        nil))                                         ; {pfx0}
\end{verbatim}

At this point, students can think about other properties
of the {\tt prefix} operator and test their expectations
with DoubleCheck. One property they might think of is
a relationship between {\tt prefix} and {\tt append}.

\begin{samepage}
\label{app-pfx}
\begin{center}
\begin{tabular}{lll}
({\tt prefix}~({\tt len} $xs$) ({\tt append}~$xs ~ ys$)) = $xs$  & ~~~~~~~~~~ & \{\emph{app-pfx}\}\\
\end{tabular}
\end{center}
\end{samepage}

A formal, DoubleCheck definition of this property would mechanize the test.
\begin{verbatim}
  (defproperty app-pfx ; preliminary version
    (xs :value (random-list-of (random-integer))
     ys :value (random-list-of (random-integer)))
    (equal (prefix (len xs) (append xs ys))
           xs))
\end{verbatim}

All of the random tests that DoubleCheck generates pass,
so the next step is a paper-and-pencil proof that the property holds for all lists.
That proof, an induction on the length of $xs$, succeeds.

Unfortunately, ACL2 fails to complete a formal proof
of the theorem corresponding to the {\tt app-pfx} property.
The problem is that, while the property holds for all of the
random tests that DoubleCheck generates, it does not hold under all circumstances.
If $xs$ is not a list, but is, instead, some other kind of object, the property fails.
To be an ACL2 theorem, the property must constrain
$xs$ to the domain of lists.\footnote{This is, of course, a vestige
of type, and we do not talk much about types.
Our students have raised questions on many topics in the course,
but none have ever asked about types, so we think our decision
to avoid the issue and use ad hoc explanations where necessary
works in the ACL2 environment, at least at the level presented in this course.}

\begin{verbatim}
  (defproperty app-pfx ; provable version
    (xs :value (random-list-of (random-integer))
     ys :value (random-list-of (random-integer)))
    (implies (true-listp xs)
             (equal (prefix (len xs) (append xs ys))
                    xs)))
\end{verbatim}

Even then, ACL2 does not succeed with a proof until it
imports the standard theorems of numeric
algebra, which have been derived in a certified package
distributed with the ACL2 system \cite{HuntKrugArith}.
So, in this example,
students must deal with a few of the complications
that can arise when moving from an informal environment
to a formal one.

One of the reasons we chose ACL2 for this course rather
than another mechanized logic, such as Isabelle \cite{isabelle},
Coq \cite{coq}, or Agda \cite{agda} is because the learning
curve for ACL2 allows students to succeed quickly
in problems of moderate complexity.
Other theorem proving systems may be equally effective.
We have not tried others in the classroom, so we are guessing
based on our own assessment of the difficulty of
applying other systems in the types of examples that we use in our course.

Our guess is based in part on the collections of
powerful theorems that come with ACL2.
Students import theorem collections when ACL2 gets stuck.
The two collections they learn to import are the ones
that imbue ACL2 with facilities for manipulating
formulas in numeric algebra. One of the collections
concerns ordinary arithmetic, the other, modular arithmetic.
These collections of theorems provide the support needed
to complete homework projects.

Examples and projects in the course progress from
simple ones like {\tt prefix} to complex ones like
merge-sort and AVL tree operations.
All of them include testing and deriving, informally and formally,
additional properties from definitional ones.
Students are not asked to create programs of this complexity from scratch,
but they are required in homework projects and on exams to extend properties
to wider domains and to develop new properties.
They are given guidelines for developing such properties \cite{PagePBtesting}.

\subsection{Propositional Logic and Digital Circuits}
\label{sec:logic/circuits}

Equations provide a basic theme that permeates the course.
Because the equations of Boolean algebra are
so much like those of numeric algebra,
which students are familiar with, the technical part of the course
starts in the domain of equations.

We derive propositional logic from ten basic equations
of Boolean algebra (Figure~\ref{fig:basic-bool-eqs}).
The traditional truth tables, absorption equations, and so on
are derived by reasoning from the basic equations in the
standard, algebraic way.

\begin{figure}
\begin{center}
\begin{tabular}{ll}
$x \vee False = x$                                   & \{$\vee$ identity\} \\
$x \vee True = True$                                 & \{$\vee$ null\} \\
$x \vee y = y \vee x$                                & \{$\vee$ commutative\} \\
$x \vee (y \vee z) = (x \vee y) \vee z$              & \{$\vee$ associative\} \\
$x \vee (y \wedge z) = (x \vee y) \wedge (x \vee z)$ & \{$\vee$ distributive\} \\
$x \rightarrow y = (\neg x) \vee y$                  & \{implication\} \\
$\neg(x \vee y) = (\neg x) \wedge (\neg y)$          & \{$\vee$ DeMorgan\} \\
$x \vee x = x$                                       & \{$\vee$ idempotent\} \\
$x \rightarrow x = True$                             & \{self-implication\} \\
$\neg(\neg x)  = x$                                  & \{double negation\} \\
\end{tabular}
\end{center}
\caption{Basic Boolean equations (axioms)}
\label{fig:basic-bool-eqs}
\end{figure}

This gives students practice, early on, in the syntax matching and
step-by-step reasoning that is used throughout the course, but
in a tightly prescribed context, where keeping track
of formulas is relatively simple.
Figure~\ref{fig:and-absorption} displays a typical proof that
students would see or be asked to derive from the basic equations,
or from other equations such as \{$\wedge$ null\} derived earlier from the basics.

\begin{figure}
\begin{center}
\begin{tabular}{lll}
    & $(x \vee y) \wedge y$                & \\
$=$ & $(x \vee y) \wedge (y \vee False)$   & \{$\vee$ identity\} \\
$=$ & $(y \vee x) \wedge (y \vee False)$   & \{$\vee$ commutative\} \\
$=$ & $y \vee (x \wedge False)$            & \{$\vee$ distributive\} \\
$=$ & $y \vee False$                       & \{$\wedge$ null\} \\
$=$ & $y$                                  & \{$\vee$ identity\} \\
\end{tabular}
\end{center}
\caption{\{$\wedge$ absorption\}: $(x \vee y) \wedge y = y$}
\label{fig:and-absorption}
\end{figure}

After some study of propositional logic,
digital circuits are introduced as an alternate notation for Boolean formulas.
We focus on and-, or-, and not-gates, but we also draw circuit diagrams with
exclusive-or, nand, nor, and other standard gate symbols.
We show, by reasoning from the basic Boolean equations,
how the behavior of any circuit can be fully realized in terms of nand-gates alone.
As an exercise, students show that an implication gate is universal in the same sense as nand.

Gradually we work up to a ripple-carry adder circuit.
All of the algebraic support for twos-complement arithmetic
is defined and justified using standard equations of numeric algebra
and, formally, by defining ACL2 operators to carry out
arithmetic on binary numerals. In this way, the students see a model
in software of a ripple-carry adder, along with a diagram of a
digital circuit for the adder at the gate level.
The fact that the operations the circuit performs
are consistent with ordinary arithmetic on numbers
is proved by induction, both informally (paper and pencil) and
formally through ACL2.
In addition, a software model for bignum addition and multiplication on binary numerals
is developed, and its operations are justified by informal and formal, mechanized proofs.

This helps students understand how physical devices can perform computations,
and that provides a basis for understanding how computers work at the circuit level.
Students acquire an understanding of how circuits do what they
do, and how engineers can know, for certain, some of the operational properties of circuits,
modulo, of course, failures in physical properties of the circuits.
We define an isomorphism between circuit diagrams and formulas in logic
and discuss what this isomorphism means about the connection between
formal logic and physical devices. We claim that digital circuits are
physical representations of logic formulas and the two are, therefore, subject to
the same sort of mathematical analysis.

The same goes for the equation-based software discussed in the course.
Stretching this to conventional software would be impractical at best,
but not theoretically impossible, since
conventional software is written in a formal language.
To mention just one of the many problems in this area,
the lack of formal semantics for conventional programming languages presents
obstacles that current research has barely begun to breach.
We discuss these issues in class, more in the honors
course than in the version of the course for computer science students.
However, students are not required to understand it deeply in either course.
Examinations and homework problems do not call for a mastery of these ideas.

\subsection{Massive-Scale Computing: Websites and User-Provided Content}
\label{sec:cassandra}

A course emphasizing the logical foundations of computing can leave
students (computing majors, especially, but other students, too)
with the erroneous impression that none of this is relevant for real-world computing.
To address this situation, students are presented with a selection of
real-world applications that they are familiar with. The applications
are chosen carefully so that they resonate with students and
elaborate on principles that have already been discussed in class.

One such application is Facebook. Students see a quick
overview of the history of web applications, with a focus on the
distinction between Web 1.0 and Web 2.0 applications.  Both of
these are dynamic web applications, and students learn how the
dynamics are performed.  In particular, they see how a webpage
consists of a template that can be ``filled in'' with information
that may come from a product catalog, for example.  The key problem
is one of searching (\emph{i.e.}, finding the relevant information to display
on this web page), and that is what varies between Web 1.0 and Web 2.0
applications.

What students
learn is that Web 2.0 applications mix content from application
producers with content from consumers.
Different consumers may see completely different results,
due to customization and personalization.
Facebook, the ultimate Web 2.0 application, actually
provides very little content in the traditional sense.
Most of the content is contributed by each user's circle of friends.

Students also learn the concept (but not the working details) of
relational databases and how they can be used to generate content for
Web 1.0 applications, such as traditional storefronts.  Students see
simple SQL queries, although they do not learn how to write their
own queries, nor how to interact with a databases.  Instead, they
see relational
databases as a solution to the key problem of finding the relevant
information to display.  Students can write the program that finds
this information in a list, and they see relational databases as
more complicated versions of this process---which is true, while
being an unmitigated oversimplification.  This reinforces one of the
big ideas in the course, namely that different definitional properties
can produce the same results but at vastly different computational
expense.

This point is driven further.  The students
then learn the Achilles heel of relational databases, namely that join
operations on large tables are prohibitively expensive. As a case in
point, we observe that the join of the ``friends'' and ``statuses'' tables that could,
in principle, support Facebook is infeasible---another big idea in
the course, that computation expense can make some devices feasible
and others infeasible.

So how does Facebook do it?  Facebook developers solved this problem
by building their own non-relational database called Cassandra~\cite{cassandra}.
At a high level, Cassandra acts like a simple key-value store, and the
students have experience with this concept, having earlier studied
AVL trees. Again, students see the big idea that different (more efficient
or scalable) solutions of the same principle can yield vastly different
computational expense.
But Cassandra is more than just a key/value store.
It features concepts such as consistent hashing, database sharding, data
replication, and eventual consistency. Although the details of these
features are very technical, at a high level of abstraction they are
accessible, even to non-technical students.
For example, students learn about Cassandra's famous ring architecture,
and they quickly grasp how this architecture enables both the massive throughput
required by Facebook and reliability in case of system failure,
which is certain to occur from time to time in such a large application.

This discussion does not prepare students to understand how NoSQL
databases such as Cassandra are implemented, or even how to use them.
Rather, it serves to convince students that ideas and
techniques that they have already seen can be scaled up to build
large, important applications such as Facebook.

To give some idea of the depth of the material on Cassandra,
it is discussed in one, 75-minute lecture, including questions
from students and a short discussion period near the end.
These discussions have been lively, which indicates that
the majority of the students get enough out of the lecture to
make intelligent observations about the topic.

\subsection{Massive-Scale Computing: Web Search Engines}
\label{sec:map-reduce}

Another example, Google's technology stack,
reinforces the lesson that the basic computer principles students
have studied in the course have real-world implications.
The students are, of course, familiar
with Google products, such as the search engine and Gmail, as paragons of web applications.
Now we expose them to a piece of Google's
technology---MapReduce~\cite{mapreduce}.
This topic is discussed in a manner similar to the discussion of Cassandra
(Section \ref{sec:cassandra}):
at a high level only in one 75-minute lecture, including a lively class discussion.

MapReduce is part of Google's approach to distributed computing.
Jobs are broken down into map and reduce steps that operate on
dictionaries (key/value data structures).  A MapReduce program
can be developed entirely on a single computer. Then, the MapReduce
framework takes care of the details of executing
individual map and reduce tasks on hundreds or thousands of
computers.

This particular technology is chosen because it is easily motivated by
Google's massive scaling needs.
In this way students see, once more, how dealing
with large scale is the main difference between computing concepts
as they have experienced them
and engineering as practiced in real-world computing.
Another reason for choosing MapReduce is because of its
roots in functional programming, which immediately ties into
programming concepts that students have learned in the course.

Although MapReduce was originally implemented in C++, its key ideas
can be rendered in many other languages, such as Java in the Apache Hadoop
project~\cite{hadoop}.  In this class, students see some MapReduce
operations expressed in the form of ACL2 functions.
They see simple examples, such as distributed word count and
distributed grep---classic examples from the MapReduce literature.

Students also see how the MapReduce framework can be used to
perform meaningful work at scale, such as inverting the graph of
internet links, which is then used to calculate
PageRank~\cite{pagerank} for Google's search results.

An important aspect of this presentation is that the (admittedly simple)
programs students see are complete.  A version of the MapReduce framework
is implemented in ACL2, so students can see how the map and reduce functions
are combined to produce a complete solution.  This solution is sequential,
and it runs on just one machine.  But students learn that the full MapReduce
framework distributes these tasks across hundreds, even thousands of computers.
Again, students see the big idea that different sets of definitional properties
(\emph{i.e.}, the sequential and distributed versions of MapReduce) produce the same
results at vastly different computational expense, and that the difference
is sufficient to make large MapReduce applications
(indexing the web, for example) feasible.

This example solution to a problem of massive scale,
another other of the big ideas in the course, together with
the Cassandra solution to accessing massive, fast changing, data bases
discussed in Section \ref{sec:cassandra}, together
provide some illumination of the massive scale idea.

\subsection{Other Big Ideas}
\label{sec:big-ideas}

One of the big ideas in the course is
how important, complex algorithms derive from simple, definitional properties.
The two most complex algorithms that we discuss in detail are merge-sort and
AVL-tree insertion. Yet, the approach to these algorithms is the same as for
all other software artifacts we present.

Namely, we assume someone has given us an operator
that carries out the computation of interest.
We look for properties that we expect the operator
would have if it worked properly, and we try to find a collection of properties
that are consistent, complete, and constructive,
since such properties must define an operator.
Unless we misconstrue our expectations,
that operator will be the one we want to define.

For example the merge portion of the merge-sort algorithm assumes
that its arguments are lists whose elements are arranged in order by increasing value.
We divide the data space into three parts:
(1)~the first argument is the empty list,
(2)~the first argument is not empty, but the second argument is, and
(3)~both arguments are non-empty.
That covers all the possibilities,
and none of the three parts of the data space
overlaps with another,
so our equations will automatically be consistent.

When either argument is empty, the result must be the other argument,
so equations corresponding to these conditions are easy to write down.
In case (3), when both arguments are non-empty, the first element of the
merged list must be the smaller of the first elements in the lists
supplied as arguments, and the rest of the elements must be the merge of the
remaining elements of list with the smaller first element
and all of the elements in the other list. Assuming the merge operator
we were given works, that analysis yields the following equations.

\begin{tabular}{lll}
~ & ({\tt merge} ({\tt cons} $x$ $xs$) ({\tt cons} $y$ $ys$)) = ({\tt cons} $x$ ({\tt merge} $xs$ ({\tt cons} $y$ $ys$))) if $x \le y$ & \{mrg$\le$\}\\
~ & ({\tt merge} ({\tt cons} $x$ $xs$) ({\tt cons} $y$ $ys$)) = ({\tt cons} $y$ ({\tt merge} ({\tt cons} $x$ $xs$) $ys$)) if $x > y$   & \{mrg$>$\}\\
\end{tabular}

As we noted earlier, the equations will be consistent because
we have divided the data space into three, non-overlapping parts.
They are constructive because, in both of the inductive references
to {\tt merge}, one of the arguments is a shorter list than
the one we started with, which makes it closer to the corresponding argument
on the left-hand side of a non-inductive equation.

The {\tt merge-sort} operator, which of course refers to the {\tt merge} operator,
is equally straightforward to define.
None of the definitions for {\tt merge-sort} and its supporting operators
is more than a few lines long.
The same is true of the AVL insertion
code, given an appropriate breakdown of the problem.
So, these complex computations derive from simple properties.

Furthermore, we can use the equations, along with a model of basic,
one-step operators such as {\tt cons}, {\tt first}, {\tt rest}, selection
(that is, {\tt if}, once the Boolean condition has been computed),
relational operators for numbers, and the like to derive
recurrence equations for the number of steps in a merge-sort computation.
We solve these equations by proposing a solution and proving it by induction,
and we find that merge-sort is an $n~log(n)$ computation.
A similar derivation of equations for insertion-sort and an analysis
of the number of computational steps these equations lead to
shows that insertion-sort is an $n^2$ algorithm.

We do not discuss in the course details in the implementation
of functional languages that affect the computational efficiency
of individual operations such as {\tt cons}, {\tt first}, and {\tt rest}
in our computational model.
Instead, we focus on another big idea in the course, namely
that different definitional properties can produce
the same results at vastly different computational expense.
A table comparing the growth of $n~log(n)$ with that of $n^2$
shows just how infeasible the insertion-sort algorithm is
for large data sets, another big idea in the course.

Thus, all the big ideas listed in Section \ref{sec:one-and-done}
are discussed in the course, but at widely different levels of detail.
There are, of course, many different ways to slice and dice the material
to be covered. For example, one could choose to cover fewer algorithms
and replace the omitted ones with more material on interactive software.
As it stands, we talk about interactive software early in the course
and discuss a model for it based on the methods employed in Racket software \cite{racket}.
But, the students neither write any interactive software nor reason about it,
so its presence in the course is a minor one.
In summary, this report describes a particular set of choices
about what details to include where, and with respect to what big ideas
in a course on computational thinking for courses in which most
students are not majoring in computer science.

\section{Related Work}
\label{sec:related-work}

This is not the first course that is designed to expose non-majors to computer science.
There is an ongoing effort by the College Board to promote a new introductory
computer science course designed for advanced high school students or first-year college
students~\cite{APcourse}.  \emph{Computer Science: Principles} (CSP), the new course initiative,
stands in contrast to the College Board's existing introductory computer science course
and Advanced Placement (AP) Computer Science exam.  The big change is that the CSP course
does not focus on programming. Instead, it spends more time on trying to develop
key insights on the nature and especially the relevance of computing in the modern world.

We share the College Board's goals, and we consider our course to be in the spirit of the
CSP approach.  However, there are significant differences between our approach and that of
the existing official pilots for CSP~\cite{csprinciples:pilots}.  Like ours, all of these
courses involve programming to some extent, but without dwelling on the programming details.
However, most of these courses use scripting languages to facilitate programming.
This makes it possible for students to
build sophisticated programs with a small amount of effort.
For instance, Python is a popular choice of language,
and so are visual languages like BYOB
(aka SNAP!) and App Inventor~\cite{byob-paper,app-inventor}.
These languages allow students to build graphical and mobile applications
like the ones they normally interact with.

For our course we chose to use the language ACL2 instead of a scripting language
because the theorem prover that comes with
ACL2 makes it easier for the students to \emph{reason} about the
programs they build.
We think the introduction of methods of ensuring software correctness
adds value to discussions about the relevance of computer software,
so we chose to include material on testing and verification
at the expense of, for example, multimedia applications.

Other courses aimed at the same audience, with ``computational thinking'' as an organizing theme, have been springing up at many universities~\cite{comp-thinking}.  Common threads in these
courses include a brief introduction to programming, societal impacts of computer science,
and discussions on the limits of computing.  The big ideas in our honors course
include most of these ideas, but our approach is a logic-based
introduction to computational thinking.

An examination of computing programs at
fifteen large universities in the central United States
suggests that about a third of such programs
require a course in logic that is separate from the required discrete math course.
In many cases, the required logic course focuses on digital circuits,
and those courses rarely include any material on reasoning about properties of circuits.
However, three of the fifteen programs did offer a course
on reasoning about circuits and/or software.

Evidence of the integration of mechanized logic into undergraduate
courses is harder to find.
Manolios uses ACL2,
via the ACL2 Sedan \cite{acl2s}, in a lower division logic course
at Northeastern University that was
introduced in an experimental form by Felleisen \cite{acl2-freshman}.
The Manolios course, which is called ``Logic and Computation,'' is probably
the work that relates most closely to this paper.
Our reading of the course description is that it covers ACL2 in greater
detail than our course, but contains less ``big picture'' material.

Harper, Pfenning, and Erdmann have revised computer science courses
at Carnegie Mellon University to incorporate reasoning about software
throughout the curriculum, right from the beginning. Erdmann's notes
\cite{erdmann} on a lower division course called ``Principles of Programming''
and Pfenning's notes \cite{pfenning} on ``Principles of Imperative Computation''
provide some insight into how Carnegie Mellon is dramatically
recasting computer science to have reasoning as a central element of
computer programming.

Tinelli \cite{tinelli} has included
assignments in the use of KeY \cite{key}, a tool for theorems and proofs
about programs written in a subset of Java. Courses employing
Haskell often use QuickCheck \cite{classenQuickCheck}, which gives students practice
in stating properties as logic formulas, an important skill in using
computational logic systems. Jackson's Alloy system is used in
undergraduate classes and exposes students to logic as a tool for
stating and verifying properties of software components \cite{alloy}.

Our review of related work is not comprehensive.
No doubt there are many efforts that we have not managed to find.
The literature in this area is sparse, but we think the papers and projects
discussed in this section provide a context in which to consider the work
reported in this paper.

\section{Where Do We Go from Here and Why?}
\label{sec:conclusion}

A course including topics in classical logic, digital circuits, programming,
testing, verification, and other major computing concepts,
all of it couched in terms of a familiar form of reasoning,
namely algebraic equations,
can provide a comfortable, yet challenging environment for interested students,
regardless of background, assuming a standard, college-prep education
that includes high-school algebra.
Such a course can provide a basis
for understanding in a fundamental way what computers do and how they do it.
It is one way to introduce students to computational thinking.

This is not a ``soft skills'' course.
It calls for careful thinking, and it rewards studious attention.
Yet, it is accessible and interesting to a broad base of college students.
This combination of depth, challenge, and reward offers students
something new and valuable.

An essential component of this enterprise is an equation-based programming
language with a property-based testing facility. A mechanized logic with
a quick-entry learning curve enhances the experience and the
educational impact relative to the learning effort that students invest.
A conventional programming language will not serve because it cannot be
understood in terms of classical logic
and the standard mechanisms of algebraic reasoning.
The theme of equations cannot be carried throughout a course
in which the programming component relies on the imperative paradigm.

We are writing a textbook that takes the approach and includes the material
discussed in this paper.
Drafts of the text have been used to provide readings in the course.
We plan to develop interactive, web-accessible learning tools,
including mini-tutorials, exercises, and automated assessment of
solutions for instant feedback. Lecture notes, homework projects,
and examinations are available upon request to educators who want to incorporate
some of these ideas into their work.

The new course in the principles of
computer science proposed by the College Board is
designed for college preparatory students and first-year
college students~\cite{APcourse}.
It emphasizes computational thinking.
The proposal elaborates seven big ideas and key concepts:
creativity, abstraction, data, algorithms, programming, internet, and impact.
Based on the descriptions in their proposal,
we believe that the material in our course
and its accompanying text provide an effective
learning environment for those ideas and concepts.
A course following this approach would be one way to introduce
computational thinking to
a broad range of students.

\subsubsection*{Acknowledgments.} The authors are grateful to the reviewers
for their thoughtful and detailed comments and suggestions.
Their ideas led to significant improvements in the paper,
especially in the level of detail describing elements of the course,
relating those to its big ideas,
and in discussing the experiences that students
reported in their assessments of the course.

\bibliographystyle{splncs}
%%\bibliography{rgrptfpie2012}

\end{document}